\documentclass[aps,prl,reprint,superscriptaddress,amsmath,amssymb]{revtex4-2}

\usepackage{graphicx}
\usepackage{listings}
\usepackage{longtable}
\usepackage{xcolor}
\usepackage{algorithm}
\usepackage[bookmarks,bookmarksnumbered,colorlinks,allcolors=blue]{hyperref}
\usepackage{bookmark}
\usepackage{ifthen}
\usepackage{multirow}
\usepackage{array}
\usepackage{siunitx}
\usepackage{qcircuit}
\usepackage[normalem]{ulem}

\newcolumntype{M}[1]{>{\centering\arraybackslash}m{#1}}

\definecolor{dkgreen}{rgb}{0,0.6,0}
\definecolor{gray}{rgb}{0.5,0.5,0.5}
\definecolor{mauve}{rgb}{0.58,0,0.82}

\newcommand{\ket}[1]{|{#1}\rangle}

\begin{document}

\author{Tenghui Wang}
\email{wthzju@gmail.com}
\affiliation{DAMO Quantum Laboratory, Alibaba Group, Hangzhou, Zhejiang 311121, China}
\author{Feng Wu}
\affiliation{DAMO Quantum Laboratory, Alibaba Group, Hangzhou, Zhejiang 311121, China}
\author{Fei Wang}
\affiliation{DAMO Quantum Laboratory, Alibaba Group, Hangzhou, Zhejiang 311121, China}
\author{Xizheng Ma}
\affiliation{DAMO Quantum Laboratory, Alibaba Group, Hangzhou, Zhejiang 311121, China}
\author{Gengyan Zhang}
\affiliation{DAMO Quantum Laboratory, Alibaba Group, Hangzhou, Zhejiang 311121, China}
\author{Jianjun Chen}
\affiliation{DAMO Quantum Laboratory, Alibaba Group, Hangzhou, Zhejiang 311121, China}
\author{Hao Deng}
\affiliation{DAMO Quantum Laboratory, Alibaba Group, Hangzhou, Zhejiang 311121, China}
\author{Ran Gao}
\affiliation{DAMO Quantum Laboratory, Alibaba Group, Hangzhou, Zhejiang 311121, China}
\author{Ruizi Hu}
\affiliation{DAMO Quantum Laboratory, Alibaba Group, Hangzhou, Zhejiang 311121, China}
\author{Lu Ma}
\affiliation{DAMO Quantum Laboratory, Alibaba Group, Hangzhou, Zhejiang 311121, China}
\author{Zhijun Song}
\affiliation{DAMO Quantum Laboratory, Alibaba Group, Hangzhou, Zhejiang 311121, China}
\author{Tian Xia}
\affiliation{DAMO Quantum Laboratory, Alibaba Group, Hangzhou, Zhejiang 311121, China}
\author{Make Ying}
\affiliation{DAMO Quantum Laboratory, Alibaba Group, Hangzhou, Zhejiang 311121, China}
\author{Huijuan Zhan}
\affiliation{DAMO Quantum Laboratory, Alibaba Group, Hangzhou, Zhejiang 311121, China}
\author{Hui-Hai Zhao}
\affiliation{DAMO Quantum Laboratory, Alibaba Group, Beijing 100102, China}
\author{Chunqing Deng}
\email{dengchunqing@gmail.com}
\affiliation{DAMO Quantum Laboratory, Alibaba Group, Hangzhou, Zhejiang 311121, China}

\title{Efficient initialization of fluxonium qubits based on auxiliary energy levels}

\begin{abstract}
Fast and high-fidelity qubit initialization is crucial for low-frequency qubits such as fluxonium, and in applications of many quantum algorithms and quantum error correction codes. In a circuit quantum electrodynamics system, the initialization is typically achieved by transferring the state between the qubit and a short-lived cavity through microwave driving, also known as the sideband cooling process in atomic system. Constrained by the selection rules from the parity symmetry of the wavefunctions, the sideband transitions are only enabled by multi-photon processes which requires multi-tone or strong driving. Leveraging the flux-tunability of fluxonium, we circumvent this limitation by breaking flux symmetry to enable an interaction between a non-computational qubit transition and the cavity excitation. 
With single-tone sideband driving, we realize  qubit initialization with a fidelity exceeding 99\% within a duration of 300~ns, robust against the variation of control parameters. Furthermore, we show that our initialization scheme has a built-in benefit in simultaneously removing the second-excited state population of the qubit, and can be easily incorporated into a large-scale fluxonium processor. 
\end{abstract}

\maketitle

The initialization of qubits is integral to quantum computing, representing one of the DiVincenzo criteria~\cite{divincenzo2000physical}. Recent studies underscore the considerable impact of both the fidelity and speed of initialization on the effectiveness of quantum error correction (QEC), particularly when frequent reset  is required following the measurement of the syndrome qubits~\cite{google2021exponential}.
Relying on the natural energy dissipation of the qubit is not only time-consuming given increasing qubit coherence times, but also ineffective for low-frequency qubits where thermal excitations can significantly impact the qubit state. As such, active qubit initialization methods have been implemented in various physical platforms for quantum computing~\cite{monroe1995resolved, Vuletic1998, jelezko2004observation, Elzerman2004}.

In the realm of superconducting quantum circuits, an active initialization can be realized by processing the outcomes of projective measurements~\cite{Johnson2012Heralded, Rist2012Initialization, Salath2018low, gebauer2020state}. However, this method necessitates quantum feedback that requires additional control sources and is ultimately limited by the feedback latency. Alternatively, initialization can be implemented by transferring the qubit state into a dissipative quantum system~\cite{Reed2010fast, mcewen2021removing, zhou2021rapid, geerlings2013demonstrating,egger2018pulsed,  Magnard2018fast, Zhang2021Universal}, such as a readout cavity. Several protocols have been proposed and demonstrated, which involve bringing the qubit and the cavity into resonance, either adiabatically~\cite{Reed2010fast, mcewen2021removing} or parametrically~\cite{zhou2021rapid}. However, these protocols require the qubit to operate at a frequency that is either close to or above the cavity frequency, which limits their application in low-frequency qubits. Alternatively, a sideband transition can be used to transfer the qubit excitation into the dissipative cavity~\cite{egger2018pulsed, Magnard2018fast, Zhang2021Universal}. To comply with the selection rules~\cite{Blais2007quantum}, two weak microwave drivings or a single strong driving is needed to activate the second order transitions when symmetry breaking is absent. More than requiring additional control resources, these microwave drivings could also introduce significant ac-Stark shift~\cite{egger2018pulsed, Magnard2018fast}, which complicates the experimental calibration and renders it highly sensitive to the control parameters.

In this work, we present an efficient initialization protocol for fluxonium qubits based on the idea of sideband cooling. As a promising candidate qubit for fault-tolerant quantum computing, fluxonium has garnered significant attention because of its remarkable coherence time~\cite{pop2014coherent, nguyen2019high,somoroff2021millisecond} and its ability to perform high-fidelity two-qubit operations~\cite{moskalenko2022high, bao2022fluxonium, huang2023quantum, Dogan2023two, ding2023high, ma2023native, zhang2023tunable}. Our protocol takes the advantage of the flux tunability and the rich, anharmonic energy level structure of fluxonium. By displacing the qubit away from its flux degeneracy position~\cite{Liu2005Optical}, we establish a strong coupling between a non-computational level of the fluxonium and its readout cavity to enable sideband transitions with a weak monochromatic drive. In addition, by adiabatically increasing the driving strength, the auxiliary level acts as a dark state, facilitating the qubit population to be directly transferred into the cavity excitation, thereby significantly enhancing the initialization efficiency. Here, we select the second-excited state as the auxiliary level and achieve ground state initialization with a fidelity exceeding 99\% within a duration of 300~ns, robust against the variation of the control parameters. We further show that our scheme can be directly combined with leakage removal on this auxiliary level, and easily extended to initializing multiple qubits through frequency multiplexing.

\begin{figure}
  \includegraphics[width=8.6cm]{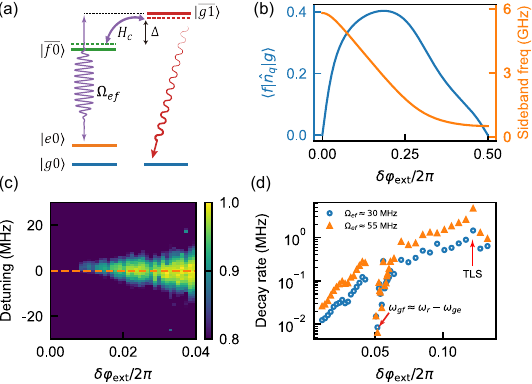}
  \caption{\label{fig1}(a) Energy diagram of the fluxonium-cavity system. The interaction facilitated by $|f0\rangle$ and $|g1\rangle$ allows the transfer of the qubit excitation in $\ket{e0}$ into the dressed state $|\overline{g1}\rangle$ through a sideband drive of strength $\Omega_{ef}$ at the frequency detuning $\Delta$. Subsequently, the population returns to the ground state, a result of strong cavity dissipation. (b) Transition matrix element of $\langle f|\hat{n}_q|g\rangle$ and the sideband frequency versus the external flux shift $\delta \varphi_{\textrm{ext}}$ away from the $\varphi_{\textrm{ext}}=\pi$ position. (c) Ground state population $P_{g0}$ (with readout error correction) versus $\delta \varphi_{\textrm{ext}}$ and the detuning respect to the sideband frequency after the application of a 30~$\mu$s drive, at $\Omega_{ef} = 55$~MHz. (d) Initialization rate versus $\delta \varphi_{\textrm{ext}}$ under two driving strengths.}
\end{figure}

The fluxonium qubit is capacitively coupled  to the readout cavity. The system is described by a coupling Hamiltonian of $H_c = \hbar g_r \hat{n}_q \hat{n}_r/2$, where $\hat{n}_{r(q)}$ denotes the Cooper-pair number operator of the cavity (qubit). The concept of our protocol is illustrated in Fig.~\ref{fig1}(a), where we label the three lowest levels of fluxonium as $|g\rangle$, $|e\rangle$, $|f\rangle$ and the $n$-photon Fock state of the cavity as $|n\rangle$, respectively. The existence of the coupling $H_c$ between the qubit and the cavity hybridizes $|f0\rangle$ and $|g1\rangle$, which are the tensor product states of the composite system. The energy eigenstate (dressed state) $|\overline{g1}\rangle$ contains the fluxonium excitation component $|f0\rangle$, enabling a population transfer from $|e0\rangle$ to $|\overline{g1}\rangle$ via red-sideband driving at the frequency $\omega_r-\omega_{ge}$ and strength $\Omega_{ef}$.
Simultaneously, the transferred population in $|\overline{g1}\rangle$ quickly relaxes to the system ground state $|g0\rangle$, due to fast photon dissipation in the cavity. We estimate that the transition rate from $|e0\rangle$ to $|g1\rangle$ is proportional to $(\Omega_{ef}g_r/{2\Delta})|\langle g|\hat{n}_q|f\rangle|$ in the dispersive regime, where $\Delta=\omega_{gf}-\omega_r$ (see Supplementary Material). 

However, at the flux degeneracy position $\varphi_{\textrm{ext}}=\pi$ which is the sweet spot for coherent qubit operations due to its insensitivity to flux noise, the potential has the parity symmetry therefore each eigenstate has well-defined even or odd parity. In particular, $|g\rangle$ and $|f\rangle$ are both even parity wavefunctions, rendering the rate of the transition $|\langle g|\hat{n}_q|f\rangle|$ to be precisely zero. To enable this direct sideband transition, we temporarily introduce a flux offset $\delta\varphi_{\text{ext}}$ to position the qubit at $\varphi_{\textrm{ext}}\ne \pi$ for breaking the parity symmetry. In Fig.~\ref{fig1}(b), we illustrate the transition matrix element $\langle g|\hat{n}_q|f\rangle$ (blue line) as well as the sideband-transition frequency $\omega_r-\omega_{ge}$ (orange line), versus $\delta \varphi_{\textrm{ext}}$. As the external flux shifted away from the $\varphi_\text{ext} = \pi$, the transition matrix element increases significantly until it reaches a maximum value, eventually becoming zero when it reaches another symmetry point at $\varphi_\textrm{ext} = 2\pi$. The calculation is based on the qubit parameters extracted from the measured qubit spectrum versus external flux $\omega_{ge}(\varphi_\text{ext})$ (see Supplementary Material).

We first demonstrate microwave activated sideband transitions enabled by symmetry breaking. Starting from the qubit operated at the sweet spot, we prepare the qubit with a $\pi/2$-pulse. A rectangular flux pulse $\delta\varphi_{\textrm{ext}}$ is then applied to shift the qubit slightly away from the sweet spot. 
Applying a fixed-strength drive for $30~\mu$s, we adjust its frequency detuning and record the ground state population $P_{g0}$ at various values of $\delta\varphi_{\textrm{ext}}$. The drive strength, $\Omega_{ef}\approx 55$~MHz, is inferred from the Rabi rate between states $|e\rangle$ and $|f\rangle$ at the sweet spot. As depicted in Fig.~\ref{fig1}(c), the transition occurs when the microwave frequency aligns with the sideband frequency $\omega_r-\omega_{ge}(\delta\varphi_{\textrm{ext}})$. As the qubit shifts away from the sweet spot with increasing $\delta\varphi_\text{ext}$, the initialization rate, indicated by the width of the measured $P_{g0}$ versus frequency detuning, increases significantly.

We characterize the initialization rate of the qubit population for a wider range of $\delta\varphi_\text{ext}$ for two specific driving strengths, $\Omega_{ef} \approx 30$ and $55$~MHz. As illustrated in Fig.~\ref{fig1}(d), the protocol functions effectively for the majority of bias points.
Notably, while a stronger drive consistently accelerates the initialization process, an increase in $\delta \varphi_{\text{ext}}$ that shifts the qubit away from the symmetry position also enhances the initialization rate. We also detect some non-monotonic features, indicated by two red arrows in Fig.~\ref{fig1}(d). The arrow on the right marks a peak in the initialization rate, signifying an acceleration of initialization due to the coupling with a dissipative two-level system~\cite{basilewitsch2017beating,Sun2023Characterization}. Conversely, at the left point where $\delta \varphi_{\textrm{ext}}/2\pi\approx0.055$, the sideband frequency $\omega_{r}-\omega_{ge}$ matches the qubit transition frequency $\omega_{gf}$, leading to a population leakage into the $|f0\rangle$ state and a consequent reduction in initialization efficiency.

\begin{figure}
 \includegraphics[width=8.6cm]{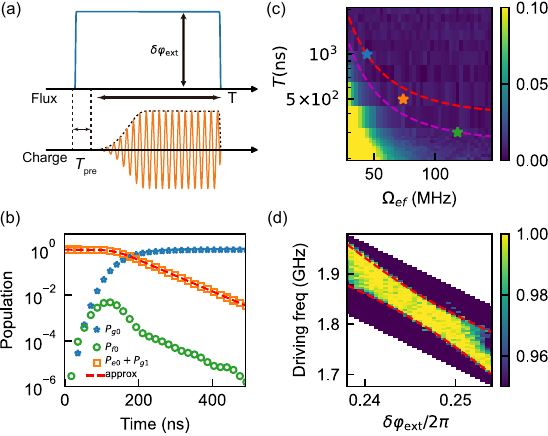}
 \caption{\label{fig3_seq}(a) Control sequence for the initialization with adiabatic state transfer. A rectangular pulse is applied in flux line to bring the qubit into the target $\delta\varphi_\text{ext}$. The orange line and black dashed line present the microwave waveform and its envelop. The driving strength slowly increase at the initial half duration before stabilizing at a fixed value for the ensuing half. The total driving duration is represented by $T$, while $T_\text{pre}$ denotes the advance duration of the flux pulse. (b) Simulation results versus the evolution time. The red dashed line represents the analytic approximation of the population $P_{e0} + P_{g1}$. (c) Initialization error versus the driving strength and duration. Two dashed lines denote the error of $10^{-2}$ and $10^{-3}$, as calculated from the simulation. Three sets of parameters marked with star are selected for statistical characterization. (d) Initialization fidelity versus the driving frequency and the flux offset with $\Omega_{ef}=71$~MHz and $T=500$~ns. Two red dashed lines represent the energy level of the two dressed states.}
\end{figure}

We further seek to improve the initialization speed by optimizing the dynamics of the system through control. While reducing the energy detuning $\Delta$ can lead to faster initialization, a small $\Delta$ combined with a large driving strength $\Omega_{ef}$ could induce population leakage to $|f0\rangle$, and limit the overall initialization efficiency.
To model the system dynamics, we rewrite the system Hamiltonian in the subspace formed by the energy levels $|e0\rangle$, $|f0\rangle$ and $|g1\rangle$ as
\begin{align}
H = \frac{1}{2}
\begin{bmatrix}
 0 & \Omega_{ef} & 0 \\
 \Omega_{ef} & 2\Delta & g_{rf} \\
 0 & g_{rf} & -i\Gamma
\end{bmatrix},
\end{align}
where $g_{rf} = g_r |\langle g|\hat{n}_g |f\rangle|$ is the effective coupling between $|g1\rangle$ and $|f0\rangle$, and $\Gamma$ is the photon emission rate of the cavity. Ignoring the non-hermitian term of $-i\Gamma$, one of instantaneous eigenstates of the subsystem
\begin{gather}
|\psi_0\rangle = \cos{\theta}|e0\rangle - \sin{\theta}|g1\rangle
\end{gather}
forms a dark state that prevents
the leakage of population to the $\ket{f0}$ state, where $\theta$ is defined as $\text{arctan}(\Omega_{ef}/g_{rf})$~\cite{giannelli2014three,torosov2014non}.
By adiabatically adjusting $\theta$, the system remains in the $|\psi_0\rangle$ state, thus maximizing state transfer at $\Delta = 0$. This subsequently facilitates the state transfer from $|e0\rangle$ to $|g1\rangle$, without necessitating the excitation of $|f0\rangle$.
The non-hermitian term $-i\Gamma$ contributes an imaginary energy $-(i\Gamma/2) \sin^2\theta$ to $\ket{\psi_0}$, leading to the relaxation of the population in both $|e0\rangle$ and $|g1\rangle$ states out of this subspace and into $\ket{g0}$ (see Supplementary Material).
Owing to the minimal non-adiabatic error to the other two eigenstates in the subspace, the total excited population can be approximated as $P_{e0}+P_{g1} \approx \exp({-\Gamma \int_0^T{\sin^2{\theta(t)}dt}})$, with $T$ representing the total evolution duration.
The time-averaged initialization rate is given by $\Gamma \langle \sin^2{\theta}\rangle$,
which increases as $\theta$ increases and is limited by the photon emission rate of the cavity.

The control scheme under discussion is depicted in Fig.~\ref{fig3_seq}(a). Throughout the sequence, a flux pulse $\delta\varphi_{\textrm{ext}}$ is utilized to align $|f0\rangle$ and $|g1\rangle$. Following a brief delay of $T_\text{pre} = 10$~ns, we gradually increase $\theta$ by increasing the microwave driving strength $\Omega_{ef}$ initially, and then sustaining it at a steady level. The ramp time is half of the duration, $T/2$. In order to minimize nonadiabatic transitions, we incorporate a pulse-shaping technique~\cite{martinis2014fast} for the envelope (see Supplemental Material).
To confirm the feasibility of this adiabatic state transfer, we initially perform a simulation with an initial state of $|e0\rangle$, selecting $\Omega_{ef} = 71$~MHz , $\Delta=0$, and $T = 500$~ns. As illustrated in Fig.~\ref{fig3_seq}(b), the total population $P_{e0} + P_{g1}$ aligns with our analytical model, and the system rapidly transitions to its ground state $|g0\rangle$. Concurrently, the leakage $P_{f0}$ remains minimal and ultimately falls below $10^{-5}$ at the end of the evolution.

In the conducted experiment, we measure the initialization error of our adiabatic state transfer protocol. The error $e_i = 1-P_{g0}$ is characterized by comparing the magnitude of the readout signal contrast followed by a Rabi oscillation after state initialization represented as $r_\text{rabi}$, and the maximal value of $|\vec{r}_g-\vec{r}_e|$. Here, $r_{\text{rabi}} = (1 - 2e_i) |{\vec{r}_g-\vec{r}_e}|$ (see Supplementary Material). The term $\vec{r}_{g(e)}$ represents the central point of the readout distribution for the ground (excited) state in the IQ plane, which can be inferred by fitting the distribution with a Gaussian~\cite{bao2022fluxonium}. 
In Fig.~\ref{fig3_seq}(c), we present the measured initialization error versus $T$ and $\Omega_{ef}$ along with contours corresponding to $10^{-2}$ and $10^{-3}$ errors estimated from the simulations. In agreement with the simulations, the measured errors display a decreasing trend as increasing driving strength and duration. According to the simulations, for a large variation of $\Omega_{ef}$, the initialization error can be reduced to below $10^{-2}$ in less than 1~$\mu s$ and can be further improved to $10^{-3}$ in 400-500~ns for $\Omega_{ef}>100$~MHz. Additionally, we repeat the measurement on three different set of parameters, $\Omega_{ef}=\{43, 71, 114\}$~MHz and $T=\{1000, 500, 300\}$~ns (marked with stars) for statistics purposes. 
The measured initialization errors are $0.62\%\pm0.24\%$, $0.66\%\pm0.19\%$, and $0.63\%\pm0.21\%$, respectively.

We also employ the measured initialization rate to estimate the lower limits of these errors when the system attains a stationary state, which are 0.072\%, 0.042\%, and 0.031\%, respectively. Detailed information regarding error statistics and estimations can be found in the Supplementary Material. These lower limits are notably smaller than our measurements. The discrepancy between the experiment and theory might be attributed to the state excitation during readout. Nevertheless, we achieved qubit state initialization with over 99\% fidelity within a vast range of $\Omega_{ef}$ and $T$. To further assess the robustness with respect to other parameters, we fix $\Omega_{ef}=71$~MHz and $T=500$~ns, and sweep the driving frequency and $\delta\varphi_{\textrm{ext}}$. The measured fidelity of the initialization are presented in Fig.~\ref{fig3_seq}(d). Within the region delineated by the two energy levels of $|\overline{f0}\rangle$ and $|\overline{g1}\rangle$ (indicated by two red dashed lines), we achieve high-fidelity initialization over a frequency span approaching 100~MHz.

\begin{figure}
 \includegraphics[width=8.6cm]{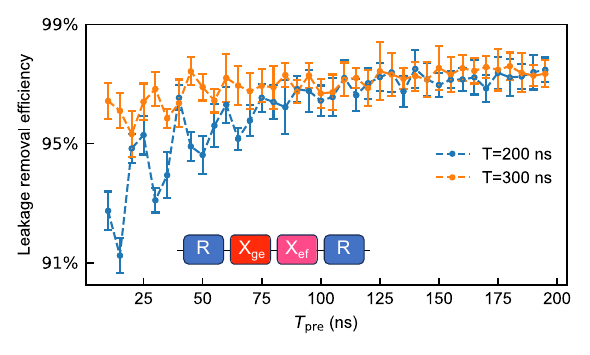}
 \caption{\label{fig3}Leakage removal efficiency versus $T_\text{pre}$ at $\Omega_{ef}=114$~MHz. The inset shows the control sequence, where $|f0\rangle$ is prepared by a qubit initialization or reset operation ($R$) followed by two $\pi$-pulses, $X_{ge}$ and $X_{ef}$.}
\end{figure}

Putting this scheme in the context of QEC, we explore its potential in addressing leakage errors and its applicability to the initialization of multiple qubits. Leakage errors, which typically accumulate with the number of gate operations, are generally hard to be detected and subsequently recovered by QEC~\cite{mcewen2021removing,GhoshUnderstanding13, suchara2015leakage}. Therefore, it is desirable to eliminate the out of computational-state excitations during qubit initialization~\cite{Battistel2021, Marques2023}. 
The strong resonant interaction between $|f0\rangle$ and $|g1\rangle$ results in the population of $|f0\rangle$ reverting to the ground state via cavity dissipation. We assess the effect of leakage removal by preparing the $\ket{f0}$ state and implementing the initialization protocol with parameters, $T_\text{pre} = 10$~ns and $\Omega_{ef}=114$~MHz. Using the same scheme to characterize initialization errors, the contrast in the detected readout signal can be represented as $r_\text{rabi} = (1-P_{f0}) |\vec{r}_g-\vec{r}_e|$, under the assumption that all initialization errors stem from the leakage population $P_{f0}$. The efficiency of leakage removal, $1-P_{f0}$, for a state with maximum leakage (prepared as $P_{f0}=1$) can be assessed. The observed efficiency for driving duration of $T=200$~ns and $T=300$~ns are 92.7\% and 96.4\%, respectively. An intuitive method to improve the efficiency of leakage removal involves extending the resonance duration $T_\text{pre}$ before the microwave drive. As depicted in Fig.~\ref{fig3}, we note damping oscillations in the efficiency relative to $T_\text{pre}$, indicative of the population exchange between $|f0\rangle$ and $|g1\rangle$. By extending $T_\text{pre}$ to approximately 100~ns, the efficiencies for both $T=200$~ns and $T=300$~ns increase to roughly 98\%. The integration of a pre-resonance duration $T_\text{pre}$ effectively eliminates the population in $|e0\rangle$ and $|f0\rangle$ with high fidelity, offering a straightforward operation for leakage removal in fluxonium qubits.

\begin{figure}
 \includegraphics[width=8.6cm]{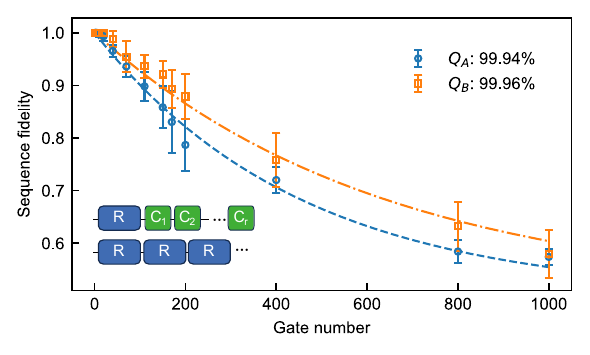}
 \caption{\label{fig4}Single-qubit gate fidelity of $Q_A$ and $Q_B$ characterized by randomized benchmarking, following the simultaneous initialization of both qubits. Each curve is individually obtained while the other qubit undergoes a repeatedly initialization process.}
\end{figure}

We ultimately illustrate the simultaneous initialization and operation of multiple qubits using this scheme. To optimize electronic resources, we employ a shared generator for the sideband driving of multiple qubits. For instance, an additional qubit ($Q_B$) utilizes the same generator as the initial qubit ($Q_A$) for the initialization, where two sideband driving tones for both qubits are generated via frequency multiplexing and broadcasted to both qubits through a power splitter, connected to both qubits' control lines (see Supplementary Material for the setup).
To validate the isolation of this initialization scheme among qubits, we employ randomized benchmarking~\cite{knill2008randomized, magesan2012characterizing} to assess the average fidelity of single-qubit gates on one qubit, while concurrently applying repeated initialization operations to another qubit. For $Q_A$ and $Q_B$, the initialization is achieved using sideband frequencies of $1.832$ GHz and $1.692$ GHz, respectively, at $\Delta=0$. The sideband driving strength and duration are set at $60$ MHz and $1~\mu$s for both qubits. The duration of all single-qubit rotations is 20~ns.
As presented in Fig.~\ref{fig4}, we find no interference with the other qubit's state initialization or single-qubit gate operations while either qubit undergoes repeatedly initialization. Both $Q_A$ and $Q_B$ display high single-qubit gate fidelity, with the average gate fidelity achieving 99.94\% and 99.96\%, respectively. These values are consistent with those observed when no initialization operation is performed on the other qubit.

In summary, we demonstrated an efficient initialization scheme for fluxonium qubits, using the sideband cooling technique. By adjusting the external flux of the fluxonium, we disrupt the parity symmetry of the energy eigenstates, which in turn enables an interaction between a non-computational qubit transition and the cavity excitation. This manipulation facilitates the direct sideband transition with single-tone microwave driving. We further improve the control by adiabatically transferring the qubit excitation to the lossy cavity state, achieving over 99\% initialization fidelity within a duration of 300~ns. Finally, we demonstrated our scheme is robust against parameter variations, capable of removing leakages, and applicable to the simultaneous operations of multiple qubits.

Our scheme offers a robust and scalable initialization protocol that can be readily incorporated into a large-scale fluxonium processor, thus constitutes an important technology for the demonstration of quantum error correction with fluxonium qubits.

\begin{acknowledgments}
We thank the broader DAMO Quantum Laboratory team for technical support.
\end{acknowledgments}

\bibliography{sample}

\end{document}


\author{Tenghui Wang}
\email{wthzju@gmail.com}
\affiliation{DAMO Quantum Laboratory, Alibaba Group, Hangzhou, Zhejiang 311121, China}
\author{Feng Wu}
\affiliation{DAMO Quantum Laboratory, Alibaba Group, Hangzhou, Zhejiang 311121, China}
\author{Fei Wang}
\affiliation{DAMO Quantum Laboratory, Alibaba Group, Hangzhou, Zhejiang 311121, China}
\author{Xizheng Ma}
\affiliation{DAMO Quantum Laboratory, Alibaba Group, Hangzhou, Zhejiang 311121, China}
\author{Gengyan Zhang}
\affiliation{DAMO Quantum Laboratory, Alibaba Group, Hangzhou, Zhejiang 311121, China}
\author{Jianjun Chen}
\affiliation{DAMO Quantum Laboratory, Alibaba Group, Hangzhou, Zhejiang 311121, China}
\author{Hao Deng}
\affiliation{DAMO Quantum Laboratory, Alibaba Group, Hangzhou, Zhejiang 311121, China}
\author{Ran Gao}
\affiliation{DAMO Quantum Laboratory, Alibaba Group, Hangzhou, Zhejiang 311121, China}
\author{Ruizi Hu}
\affiliation{DAMO Quantum Laboratory, Alibaba Group, Hangzhou, Zhejiang 311121, China}
\author{Lu Ma}
\affiliation{DAMO Quantum Laboratory, Alibaba Group, Hangzhou, Zhejiang 311121, China}
\author{Zhijun Song}
\affiliation{DAMO Quantum Laboratory, Alibaba Group, Hangzhou, Zhejiang 311121, China}
\author{Tian Xia}
\affiliation{DAMO Quantum Laboratory, Alibaba Group, Hangzhou, Zhejiang 311121, China}
\author{Make Ying}
\affiliation{DAMO Quantum Laboratory, Alibaba Group, Hangzhou, Zhejiang 311121, China}
\author{Huijuan Zhan}
\affiliation{DAMO Quantum Laboratory, Alibaba Group, Hangzhou, Zhejiang 311121, China}
\author{Hui-Hai Zhao}
\affiliation{DAMO Quantum Laboratory, Alibaba Group, Beijing 100102, China}
\author{Chunqing Deng}
\email{dengchunqing@gmail.com}
\affiliation{DAMO Quantum Laboratory, Alibaba Group, Hangzhou, Zhejiang 311121, China}

\title{Supplementary Material for \\``Efficient initialization of fluxonium qubits based on auxiliary energy levels"}

\maketitle

\section{Model}
\subsection{System Hamiltonian}
In the subspace constructed by $|e0\rangle$, $|f0\rangle$, and $|g1\rangle$, the Hamiltonian within the laboratory frame, with the voltage drive $H_d = V_d(t) \cos(\omega_p t) {\hat{n}_q}$ applied between $|e0\rangle$ and $|f0\rangle$, is represented as
\begin{equation}
H(t) = 
\begin{bmatrix}
 \omega_{ge} & \Omega_{ef} \cos(\omega_p t) & g_{re}/2 \\
\Omega_{ef} \cos(\omega_p t) & \omega_{gf} & g_{rf}/2 \\
 g_{re}/2 & g_{rf}/2 & \omega_r-i\Gamma/2
\end{bmatrix},
\end{equation}
where $\Omega_{ef} = \langle e0|V_d(t) \hat{n}_q|f0\rangle$ and $\omega_p$ represent the driving strength and driving frequency, respectively. 
Given a particular driving voltage, the driving strength is proportionally to $|\langle e| \hat{n}_q |f\rangle|$, whereas $g_{rl}$ is equivalent to $g_r|\langle g| \hat{n}_q |l\rangle|$ ($l = e \text{ or } f$). The non-Hermitian term $-i\Gamma$ signifies the process of photon emission. Within the rotating frame $R(t) = e^{i\omega_{ge} t}|e0\rangle \langle e0| + e^{i(\omega_{ge}+\omega_p)t} |f0\rangle \langle f0|+e^{i(\omega_{ge}+\omega_p)t}|g1\rangle \langle g1|$, under the assumption that $\Omega_{ef},\,g_{re}\ll\omega_p$ and employing the rotating wave approximation, the system Hamiltonian can be simplified to
\begin{equation}
H_R = R(t) H R^\dagger(t)  + i \dot{R}(t) R^\dagger(t) =
\begin{bmatrix}
0 & \Omega_{ef}/2  & 0 \\
\Omega_{ef}/2 & \omega_{ef}-\omega_{p} & g_{rf}/2 \\
 0 & g_{rf}/2 & \omega_s-\omega_{p} -i\Gamma/2
\end{bmatrix},
\end{equation}
where $\omega_{ef}=\omega_{gf}-\omega_{ge}$ is the transition frequency between $|e\rangle$ and $|f\rangle$, and $\omega_s=\omega_r-\omega_{ge}$ is the sideband transition frequency.
In our protocol, the driving frequency is set to the sideband transition frequency, denoted as $\omega_{p}=\omega_{s}$. As a result, the detuning $\Delta = \omega_{gf}-\omega_{r} = \omega_{ef}-\omega_{p}$.
When the qubit is biased near the flux degeneracy position or sweet spot at $\varphi_\text{ext} = \pi$, $\Delta$ is around one gigahertz which is much larger than the coupling strength $g_{rf}$. We diagonalize the subspace construed by $|f0\rangle$ and $|g1\rangle$, and according to the first order perturbation theory, the dressed state $ |\overline{g1}\rangle$ can be expressed as $ |\overline{g1}\rangle\approx \frac{g_{rf}}{2\Delta} |f0\rangle + |g1\rangle$. The population transition rate to $|\overline{g1}\rangle$ is related to the transition matrix element 
$\langle e0 | H_d |\overline{g1}\rangle = \frac{\Omega_{ef}g_r} {2\Delta} |\langle g| \hat{n}_q |f\rangle|$. At the sweet spot, the transition rate is precisely zero, forbidding such direct sideband transition. The selection rule is circumvented by disrupting the parity symmetry of the wavefunctions. This is achieved by introducing a flux offset that shifts the qubit from its flux degeneracy position, as elaborated in the main text.

\subsection{Evolution}
When the qubit is biased near the resonance point, where $\Delta \approx 0$, the first order perturbation theory fails. However, by disregarding the non-hermitian term $-i\Gamma$, the reduced Hamiltonian resembles the process of stimulated Raman adiabatic passage~\cite{PetrColloquium07}. Consequently, the instantaneous eigenstates of this subspace can be derived as
\begin{align}
|\psi_+\rangle &=  \sin\theta\sin\phi|e0\rangle + \cos\phi|f0\rangle + \cos\theta\sin\phi|g1\rangle, \notag \\
|\psi_0\rangle &= \cos{\theta}|e0\rangle - \sin{\theta}|g1\rangle , \\
|\psi_-\rangle &= \sin\theta\cos\phi|e0\rangle -\sin\phi |f0\rangle + \cos\theta\cos\phi|g1\rangle, \notag
\end{align}
where
\begin{gather}
\tan\theta=\frac{\Omega_{ef}}{g_{rf}}, \notag \\
\tan\phi=\frac{\sqrt{g_{rf}^2+\Omega_{ef}^2}}{\sqrt{\Delta^2+g_{rf}^2+\Omega_{ef}^2}+\Delta}.
\end{gather}
The eigenstate $|\psi_0\rangle$ constitutes a dark state, incapable of emitting photons to $|f0\rangle$. An adiabatic increase in $\theta$ would essentially transfer the population from $\ket{e0}$ into a cavity photon $\ket{g1}$, bypassing the excitation of $|f0\rangle$. To understand this process, we perform a transformation to the spontaneous energy eigenstate basis, characterized by $H_A = R_A^{-1}H_R R_A -i R_A^{-1}\dot{R}_A$, and
\begin{equation}
R_A = \begin{bmatrix}
\sin\theta\sin\phi & \cos\theta  & \sin\theta\cos\phi \\
 \cos\phi & 0 & -\sin\phi \\
 \cos\theta\sin\phi & -\sin\theta& \cos\theta\cos\phi
\end{bmatrix}.
\end{equation}
To simplify the calculation, we take the resonant condition ($\Delta=0$), thereby setting $\phi=\pi/4$. This results in the following non-Hermitian Hamiltonian for the system dynamics:
\begin{equation}
\label{eq_adia}
H_A = \frac{1}{\sqrt{2}}
\begin{bmatrix}
\frac{\Omega}{\sqrt{2}} & i\dot{\theta} & 0 \\
-i \dot{\theta} & 0 & -i\dot{\theta}\\
0 & i\dot{\theta} & -\frac{\Omega}{\sqrt{2}} 
\end{bmatrix}  
+
\begin{bmatrix}
-\frac{1}{4} i \Gamma \cos^2{\theta}  & \frac{1}{4\sqrt{2}} i \Gamma \sin{(2\theta)} & -\frac{1}{4} i \Gamma \cos^2{\theta}  \\
\frac{1}{4\sqrt{2}} i \Gamma \sin{(2\theta)} & -\frac{1}{2} i \Gamma \sin^2{\theta} & \frac{1}{4\sqrt{2}} i \Gamma \sin{(2\theta)} \\
-\frac{1}{4} i \Gamma \cos^2{\theta} & \frac{1}{4\sqrt{2}} i \Gamma  \sin{(2\theta)} & -\frac{1}{4} i \Gamma \cos^2{\theta}
\end{bmatrix},
\end{equation}
with $\Omega=\sqrt{\Omega_{ef}^2+g^2_{rf}}$. The first matrix describes the non-adiabatic transition rate from $|\psi_0\rangle$ to $|\psi_{\pm}\rangle$, a rate directly proportional to $\dot{\theta}$. In the absence of the non-Hermitian term, the leakage error stands at
\begin{equation}
    P_\text{{leak}}^{\pm}=\frac{1}{4} \left |\int {\dot{\theta} \exp\left[-i\int{\pm \frac{\Omega(t^\prime)}{2}dt^\prime}\right]dt}\right |^2.
\end{equation}

The second matrix of Eq.~\ref{eq_adia} describes the energy decay process precipitated by the non-Hermitian Hamiltonian. The leakage and decay rates are $\frac{1}{4\sqrt{2}}\Gamma\sin{2\theta}$ and $\frac{1}{4} \Gamma\cos^2{\theta}$ respectively. When $\tan{\theta} < \sqrt{2}/2$, the decay rate exceeds the leakage rate. The leakage, instigated by the non-Hermitian Hamiltonian, is eliminated through energy dissipation. 
Theoretically, optimal control could be utilized to expedite this initialization process. However, we strategically selected a relatively straightforward control waveform to effectively demonstrate the concept. To minimize the leakage error induced by $\dot{\theta}$, we employ a series of cosine functions to represent the waveform amplitude, $V_d(t) = V_0 \tan{\theta_p(t)}$ with $\theta_p(t) = \frac{2\theta_f}{T} \sum_{n=1}^N \lambda_n (t-\frac{T}{4n\pi}\sin(4n\pi t/T))$~\cite{martinis2014fast}. 
Here, $V_0$ stands for the target voltage of the microwave drive, $\theta_f$ is the final value at the end of the pulse, fixed at $\pi/4$ in our experiment , and $\lambda_n$ denotes the coefficients of the series with $\lambda = \{1.028,-0.0606, 0.0052, 0.0055, 0.0047, 0.0046, 0.0035\}$. After the ramp up stage, the amplitude maintains the same for the rest of the driving duration, $T/2$. Hence, the total microwave amplitude can be represented by 
\begin{equation}
V_d(t)= \begin{cases}
V_0 \tan(\theta_p(t)) ,\quad &(0 \le t < T/2) \\
V_0 \tan(\theta_f),\quad &(T/2 < t \le T).
\end{cases} 
\end{equation}
It should be emphasized that $\theta_p$ equals to $\theta$ only if $g_{rf}$ equals to $\Omega_0=\langle e0|V_0 \hat{n}_q|f0\rangle$. For the convenience of the experiment, we did not deliberately align these two parameters. Because of the negligible leakage from $|\psi_0\rangle$ to $|\psi_{\pm}\rangle$, the population of $|\psi_0\rangle$ can be estimated as
\begin{equation}
    P_{\psi_0} \approx P_{g1} +P_{e0} = \exp\left(-\Gamma\int{\sin^2{\theta}}dt \right). \label{eq_init_error}
\end{equation}

\begin{figure}
  \includegraphics[scale=1.0]{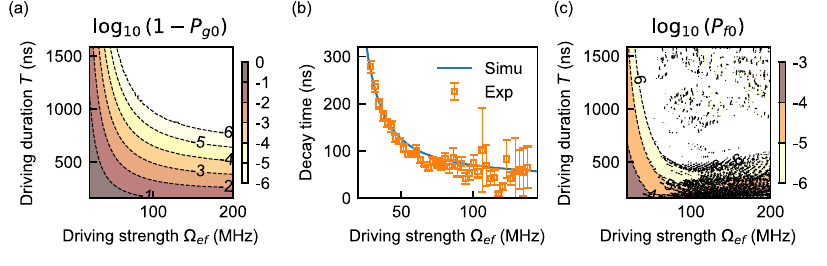}
  \caption{\label{simulation}(a) Initialization error $1-P_{g0}$ from simulation as a function of the driving duration $T$ and strength $\Omega_{ef}$. (b) Characteristic decay times of the initialization error, extracted from the simulation (solid line) and experimental data (rectangle marks), versus the driving strength. (c) Leakage error $P_{f0}$ obtained from the simulation as a function of the driving duration and strength.}
\end{figure}

We conduct numerical simulations to corroborate the previously stated analysis. The system is initially prepared in the $|e0\rangle$ state with $\Delta=0$. Fig.~\ref{simulation} illustrates the total final population of all excited states, ($1-P_{g0}$), which is presented on a logarithmic scale, for a range of driving strength and duration combinations. The initialization error, defined as $1-P_{g0}$, shows an exponential decrease with the increasing driving duration $T$ at a constant driving strength, in consistent with Eq.~\ref{eq_init_error} and presented in Fig.~2(b) of the main text. 
In Fig.~\ref{simulation}(b), we plot the characteristic decay time of the initialization error obtained from the simulation and experiment. For the experimental data, the linear mapping between the driving strength $\Omega_{ef}$ and the driving voltage is obtained by minimizing the difference between the experimental and simulated initialization rate. With the linear mapping, we find an excellent agreement between the experiment and simulation.
Indeed, faster initialization requires large driving strength, however the speed of the initialization is ultimately limited by the cavity decay rate $\Gamma$.
In Fig.~\ref{simulation}(c), we show the computed population leakage to $|f0\rangle$, which is less than $10^{-3}$ for all parameter combinations.

\section{Experimental setup}
\subsection{Device parameters}
The parameters of two fluxonium qubits, $Q_A$ and $Q_B$, are summarised in Table~\ref{table:para}. 

\begin{table}[h]
\caption{\label{table:para}Device parameters of $Q_A$ and $Q_B$.}
\centering
\begin{tabular}{M{3em} M{3em} M{3em} M{3em} M{3em} M{3em} M{3em}  M{3em} M{3em} M{3em} M{3em} M{3em}}
\hline \hline 
\multirow{2}{*}{ Qubit } & $E_{C}/h$ & $E_{L}/h$ & $E_{J}/h$ & $\omega_{ge}/2\pi$ & $\omega_{ef}/2\pi$ & $\omega_{r}/2\pi$ & $2\pi/\Gamma$ & $g_{rf}/2\pi$ & $T_1$ & $T_2^{echo}$ \\
&(GHz) & (GHz) & (GHz) & (GHz) & (GHz) & (GHz) & (ns) & (MHz) & ($\mu$s) & ($\mu$s)\\
\hline
$A$ & $1.531$ & $0.685$ & $4.164$ & $0.696$ & $4.322$ & $6.503$ & $40$ & $56$ & $35.6$ & $34.9$\\
$B$ & $1.524$ & $0.693$ & $4.275$ & $0.672$ & $4.374$ & $6.686$ & $59$ & $77$ & $30.5$ & $17.5$\\
\hline \hline
\end{tabular}
\end{table}

\subsection{Cryogenic setup}

\begin{figure}
  \includegraphics[scale=1.4]{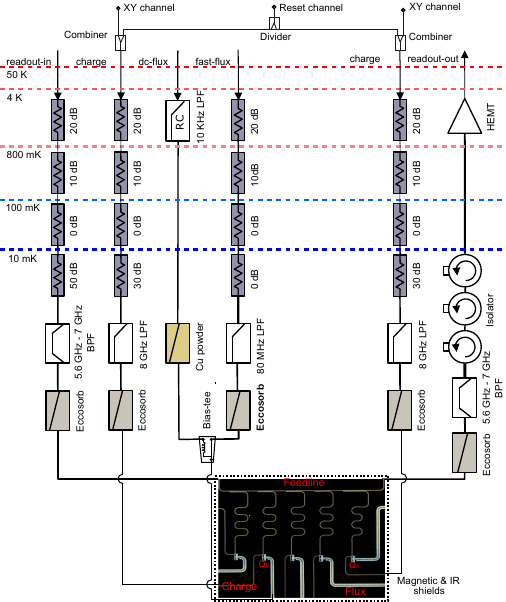}
  \caption{\label{cryogenic_setup}Wiring diagram in the dilution refrigerator. Bottom is the optical image of a processor that contains 5 fluxonium qubits. The two qubits in use in our experiment are labeled as $Q_A$ and $Q_B$.}
\end{figure}

As depicted in Fig.~\ref{cryogenic_setup}, the device is mounted in the mixing chamber plate of a BlueFors XLD-1000 dilution refrigerator, with a base temperature of approximately $10$~mK. The device is magnetically shielded with both a superconducting and mu-metal enclosure. For each fluxonium qubit, all control signals undergo substantial filtration and attenuation before reaching the devices. 
A static bias is supplied by a voltage source via the dc-flux line, maintaining the fluxonium qubit near its flux degeneracy point $\varphi_{\textrm{ext}} = \pi$. The fast flux modulation $\delta\varphi_{\textrm{ext}}$ is generated by an arbitrary waveform generator. Both signals are combined and fed to the qubit's flux control. The qubit rotation and sideband driving are applied through a common charge line. To efficiently utilize hardware resources, we divide the sideband driving from the Reset channel into two branches and merge them with the qubit rotation control signal from the XY channel. The resulting combined signal is then applied simultaneously to both qubits for implementing the initialization and single-qubit gate.
For qubit readout, the output signal is amplified by a high electron mobility transistor (HEMT) amplifier at the $4$~K stage. The wiring setup and an optical image of the device are shown in Fig.~\ref{cryogenic_setup}. 

\subsection{Fabrication}
The processor is fabricated on a 2-inch c-cut sapphire wafer, with a 100~nm thick TiN film deposited on the substrate using a reactive sputtering technique. Subsequently, it is spin-coated with 1~$\mu$m S1813 photoresist. All circuit elements, except for the Josephson junctions, are patterned by a direct laser writing system. Development is carried out in MIF-319 for 1 minute, followed by a DI water rinse. The TiN film is then etched in a RIE system using BCl3 gas, and the photoresist is removed in a sequence of acetone and IPA baths with ultrasonication. After patterning the TiN base layer, the Al/AlOx/Al Josephson junctions are fabricated using Manhattan-style shadow evaporation. The wafer is spin-coated with a bilayer e-beam resist, consisting of 800~nm PMMA A7 and 250~nm MMA EL9, and patterned in a JOEL JBX8100FS e-beam lithography system. The resist is developed in a MIBK/IPA 1:3 bath for 2 minutes, followed by a 2-minute IPA rinse. The patterned wafer is then loaded into an ultra-high vacuum cluster deposition system for the following junction formation steps: 1.1-minute ion mill for removing resist residue and native oxide, ensuring a galvanic contact between the Josephson Junctions and capacitor pads. 2. E-beam evaporation of a 40~nm base electrode Al layer, followed by a 40-minute RT oxidation at 6~Torr to form the barrier layer. 3. E-beam evaporation of a 60~nm counter electrode Al layer, followed by a 20-minute RT oxidation at 20~Torr. Finally, the lift-off process is carried out in an acetone bath for 40 minutes, followed by an acetone and IPA bath, each with 5 minutes of ultrasonication.

\section{Device calibration}
\subsection{Coupling strength and photon emission rate}
\begin{figure}
  \includegraphics[scale=1.0]{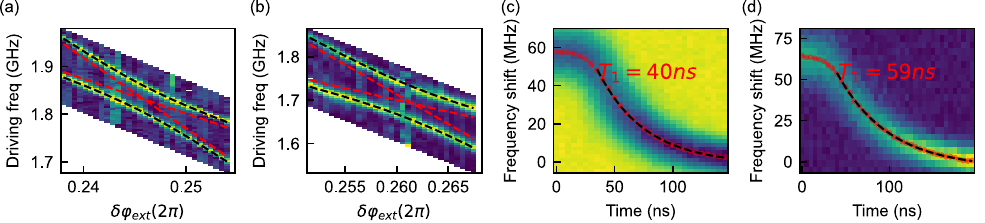}
  \caption{\label{spectrum}(a) and (b) The measured avoided level crossing between $|f0\rangle$ and $|g1\rangle$ for $Q_A$ and $Q_B$, respectively. (c) and (d) Detected qubit frequency as the function of the delay time between the microwave driving and the $\pi$-pulse for $Q_A$ and $Q_B$, respectively.}
\end{figure}

We present the characterization of the essential experimental parameters, which include the qubit-resonator coupling strength ($g_{rf}$) and the photon emission rate ($\Gamma$). We measure the coupling strength using spectroscopy. Near the resonance point, the dressed states $|\overline{f0}\rangle$ and $|\overline{g1}\rangle$ can be determined by diagonalizing the subspace of $|f0\rangle$ and $|g1\rangle$, leading to
\begin{equation}
    \label{crossing_function}
    E_{\pm}/\hbar = \frac{1}{2}(\omega_{ef}+\omega_s) \pm \frac{1}{2}\sqrt{(\omega_{ef}-\omega_s)^2 + g_{rf}^2}.
\end{equation}
In conducting the spectroscopy, we initially prepare the system in $|e0\rangle$, then measure the qubit's response to both the sweeping driving frequency and the external flux offset. The measured spectra reveal an avoided level crossing, corresponding to the transitions from $|e0\rangle$ to the dressed states $|\overline{f0}\rangle$ and $|\overline{g1}\rangle$. Within a narrow range of external flux, the bare qubit-level frequencies, $\omega_{ef}$ and $\omega_{ge}$, can be considered to linearly depend on $\delta \varphi_{\textrm{ext}}$. By substituting $\omega_{ef}$ and $\omega_s$ with linear functions of $\delta \varphi_{\textrm{ext}}$, we can use Eq.~\ref{crossing_function} to fit the spectra. The fitted dressed levels are represented by two black dashed lines, while the bare levels in the absence of coupling are plotted as two red dashed lines, as shown in Fig.~\ref{spectrum}(a) and (b). The minimum energy gap in the avoided crossing corresponds to $g_{rf}$. We find that the values of $g_{rf}/2\pi$ for $Q_A$ and $Q_B$ are $56$~MHz and $77$~MHz, respectively. At the resonance point, the corresponding flux bias and sideband frequency of $Q_{A(B)}$ are $\delta \varphi_{\text{ext}}/2\pi=0.246 (0.262)$ and $\omega_s /2\pi = 1.832 (1.692)$~GHz, respectively.

We next proceed to measure the photon emission rate of the cavity. In the dispersive regime, the photon within the cavity induces an ac-Stark shift to the qubit frequency~\cite{SchusteracStark05}. Initially, we apply a microwave driving to the cavity to stabilize its internal photon number. Followed by a idle period for photon decay after the end of the cavity driving, we apply a qubit control pulse with a variable frequency before qubit readout. This qubit control pulse corresponds to a $\pi$ pulse when its frequency matches the qubit frequency, acting as a qubit frequency probe. Given that the qubit's ac Stark shift is proportional to the photon number in the cavity, we can determine the photon relaxation time by fitting the measured qubit frequency variation versus the idle time~\cite{JeffreyFast14}. As demonstrated in Fig.~\ref{spectrum}(c) and (d), the characteristic photon emission times for $Q_A$ and $Q_B$ are determined to be $40$~ns and $59$~ns, respectively. This is achieved by fitting the decaying signal with an exponential function.

\subsection{Qubit spectrum}
The frequency of the qubit, $f_{ge}$, as a function of the external flux needs to be characterized for the extraction of qubit parameters. Such measurement is usually conducted through measuring the qubit's response to microwave driving with sweeping frequency, at different external flux positions. However, this method is often time-consuming and highly depends on the initial guess of the circuit parameters. Here, we propose an alternative protocol based on the accumulated additional dynamical phase introducing by changing the external flux. 

\begin{figure}
  \includegraphics[scale=1.0]{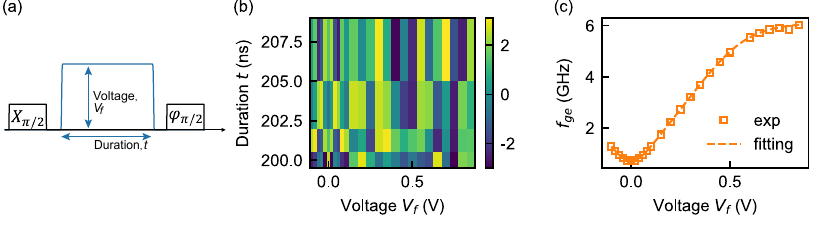}
  \caption{\label{qubit_spec}(a) Pulse sequence of the phase-based qubit spectrum measurement. (b) The measured qubit phase $\varphi_{sq}$ versus the voltage and duration of the flux pulse. (c) The extracted qubit frequency $f_{ge}$ (square point) and a fit to the qubit Hamiltonian model (dashed line).}
\end{figure}

In Fig.~\ref{qubit_spec}(a), we present the corresponding pulse sequence used in this protocol. Initially, a superposition state between $|g\rangle$ and $|e\rangle$ is prepared using a $X_{\pi/2}$-pulse, following with a rectangular flux pulse to modify the qubit frequency. The voltage and duration of the flux pulse are symbolized by $V_f$ and $t$, respectively. During the flux pulse, the qubit acquires an additional phase of $\varphi_{sq} = 2\pi [f_{ge}(0) - f_{ge}(V_f)]t + \varphi_0$, where $\varphi_0$ is a fitting parameter. Prior to the qubit readout, a second $\varphi_{\pi/2}$-pulse is applied to project the qubit state to the z-axis with an additional phase  $\varphi=(0, -\frac{2\pi}{3}, \frac{2\pi}{3})$. Assuming the measured ground state population is linearly proportional to $\langle\sigma_z\rangle$, that is, $P_g=a_0-a_1\langle\sigma_z\rangle$, and $\langle\sigma_z\rangle=\cos(\varphi_{sq}+\varphi)$, we can formulate the following equations for the three projection angles:
\begin{equation} \label{eq_P}
    \left\{\begin{aligned}
P_1&=a_0 - a_1\cos\varphi_{sq},\\
P_2&=a_0-a_1\cos\left(\varphi_{sq}-\frac{2\pi}{3}\right),\\
P_3&=a_0-a_1\cos\left(\varphi_{sq}+\frac{2\pi}{3}\right).\\
\end{aligned}\right.
\end{equation}
The parameter $a_0$ solely relies on the qubit readout, remaining invariant throughout the experiment. The parameter $a_1$ denotes the envelope of a Ramsey-type experiment, thereby reflecting the qubit's coherence time $T_2$, as represented by the equation $a_{1}(t) = a_{1,\text{max}}e^{-t/T_2}$. However, with a focus on determining the qubit phase $\varphi_{sq}$, we solve Eqs.~\ref{eq_P} for $a_0, a_1, \varphi_{sq}$ and derive
\begin{gather}
\label{eq_three_angle}
\left\{\begin{aligned}
&a_0=\left(P_1+P_2+P_3\right)/3,\\
&a_1=\sqrt{\frac{2}{3}\left[(P_1-a_0)^2+(P_2-a_0)^2+(P_3-a_0)^2\right]},\\
&\cos\varphi_{sq}=\frac{a_0-P_1}{a_1},\\
&\sin\varphi_{sq}=\frac{P_2-P_3}{\sqrt{3}a_1}.
\end{aligned}\right.
\end{gather}

Given that $a_0$ can be gauged prior to the experiment, employing two projection angles $\varphi=(0, \frac{\pi}{2})$ is sufficient to determine $a_1$ and $\varphi_{sq}$. Nonetheless, to combat the effect of measurement fluctuations, we persist in utilizing three projection angles. In Fig.~\ref{qubit_spec}(b), we present the measured $\varphi_{sq}$ of $Q_A$ in relation to the flux pulse voltage amplitude $V_f$ and the duration $t$. Considering the aliasing from sampling the phase signal, the extracted frequency variation is a multi-valued function of an integer $N$: $\Delta f_{ge}(V_f,0) = f_{ge}(V_f) - f_{ge}(0) = N f_s -\Delta \varphi_{sq}f_s / 2\pi$, where $f_s = 1 /\Delta t$ represents the sampling rate and $\Delta t$ is the minimum time step used in experiment. The term $\Delta \varphi_{sq}$ corresponds to the accumulated phase during the minimum time step. In order to restrict the value of $N$, an appropriate experimental parameters should be selected to ensure $|\Delta f_{ge}(V_f,0)|$ less than $f_s$. The implementation of this scheme can be summarized with the subsequent procedures:
\begin{enumerate}[label=(\roman*)]
    \item Given a list of pulse amplitudes, for $i$-th amplitude $V_{f, i}$, we minimize $\sum_t {\abs{\text{mod}(\varphi_{sq}(t) + 2\pi \Delta f_{ge}(V_{f,i},0) t - \varphi_0, 2\pi)}^2}$ to obtain the corresponding estimated change in qubit frequency $\Delta f_{ge}(V_{f,i},0)$.
    \item The actual qubit frequency at a given flux pulse amplitude $V_{f,i+1}$ is related to the last measured frequency $f_{ge}(V_{f,i})$, as $f_{ge}(V_{f,i+1})-f_{ge}(V_{f,i}) = \Delta f_{ge}(V_{f,i+1},V_{f,i}) + N f_s$. By setting $\Delta t = 1$~ns and selecting the appropriate the list of $V_{f,i}$, we ensure all the nearby flux pulse amplitudes satisfy $|f_{ge}(V_{f,i+1}) - f_{ge}(V_{f,i}))| < f_s$. This allows us to directly determine $\Delta f_{ge}(V_{f,i+1},V_{f,i})$ without aliasing ($N=0$) between $\Delta f_{ge}(V_{f,i},0)$ and $\Delta f_{ge}(V_{f,i+1},0)$.
    With a predetermined value of $f_{ge}(V_{f,0}=0)$, we proceed to calculate the qubit frequency with cumulative summation of frequency changes $f_{ge}(V_{f,i+1}) = f_{ge}(V_{f,0}=0)) + \sum_{j=0}^i\Delta f_{ge}(V_{f,j+1},V_{f,j})$.
    \item We convert the applied voltage into the external flux, utilizing a conversion factor obtained from previous experiments.
    \item Finally, we fit the obtained $f_{ge}$ as a function of external flux with the spectrum of the fluxonium Hamiltonian.
\end{enumerate}

The extracted $f_{ge}$, plotted in Fig.~\ref{qubit_spec}(c), aligns accurately with the fitted model. This method is highly efficient, as in principle, the qubit spectrum can be mapped out by merely selecting two time steps and a limited number of flux pulse voltages while requires little prior knowledge of the qubit parameters.

\section{metrology of initialization error}
\begin{figure}
  \includegraphics[scale=1.0]{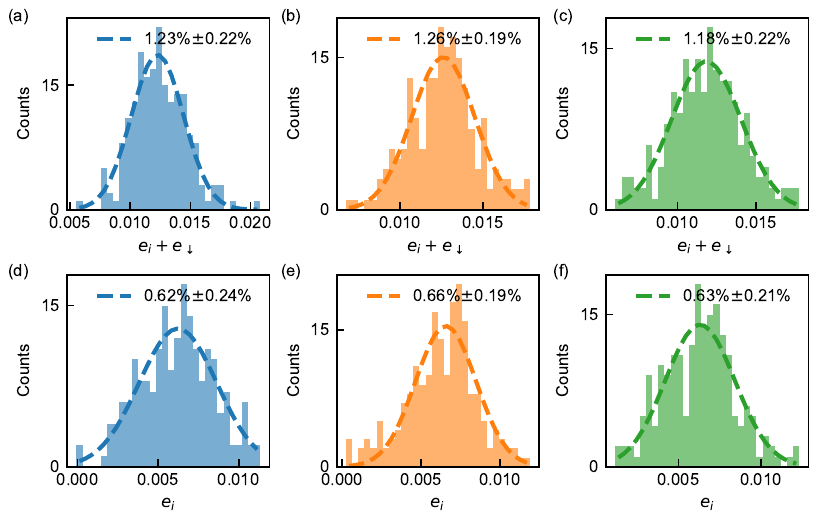}
  \caption{\label{fine_fidelity}(a-c) Histogram of $e_i+e_\downarrow$ for the parameter sets $\Omega_{ef}= \{43,71,114\}$~MHz and $T=\{1000,500,300\}$~ns. (d-f) Histogram of $e_i$ for the same sets of parameters.}
\end{figure}

The residual population of the excited state is typically estimated using a Rabi oscillation with the $|f\rangle$ state. The basic principle of this approach is given by $P_e= A_{ef} / (A_{ef} + A_{gf})$~\cite{jin2015thermal}, where $P_{g(e)}$ represents the population of the ground (excited) state, $A_{lk}$ symbolizes the detected peak-to-peak signal of the Rabi oscillation between level $|l\rangle$ and $|k\rangle$. Nonetheless, our fluxonium's dispersive readout isn't optimized for detecting the $|f\rangle$ state. The ratio of the maximum peak-to-peak amplitude of $A_{ef}$ and $A_{gf}$ to the readout noise level is not sufficient, leading to a larger uncertainty in the initialization error. 
To circumvent this uncertainty, we adopt a protocol that involves only qubit operations. Within the qubit computational space, the peak-to-peak signal of Rabi oscillation is proportional to the purity of the qubit, expressed as $A_{ge}\propto |1 - 2e_r|$, where $e_r$ denotes the initialization error preceding the Rabi oscillation. To evaluate this initialization error, we need to extract the maximal and real peak-to-peak signal of Rabi. This is accomplished by preparing an initial ground and excited state with the reset gate, then implementing the single-shot readout for each prepared state. All detected signals yield a distribution in the IQ-plane, which can be fitted with a double-Gaussian distribution given by $\sum_i^{g,e}a_i \exp(-(\vec{r}-\vec{r_i})/2\sigma^2)$~\cite{bao2022fluxonium}. By fitting the Gaussian distribution, we can extract the center point $\vec{r}_{g(e)}$ of the ground (excited) state. 
In our experiment, the average center points of each prepared state after initialization can be expressed with
\begin{align}
    \label{error_func}
    \avg{\vec{r}_{g}} &= (1-e_i)\vec{r}_g + P_{e} \vec{r}_e + P_{f} \vec{r}_f,  \notag\\
    \avg{\vec{r}_{e}} &= (P_{e} + e_\downarrow)\vec{r}_g + (1 - e_{i}-e_\downarrow) \vec{r}_e + P_{f} \vec{r}_f.
\end{align}
The total initialization error, denoted as $e_i$, can be expressed as the sum of the residual populations of the first- ($P_{e}$) and second-excited states ($P_{f}$), i.e., $e_i=P_{e}+P_{f}$. Additionally, $e_\downarrow$ represents the extra relaxation error introduced to the first-excited state during the error detection process. We note that since the excitation during the detection process is counted in $P_e$ thus in $e_i$, the experimentally obtained $e_i$ represents an upper bound of the initialization error. As the leakage in our scheme is minimal, under most cases, $e_i$ can be approximated by $P_{e}$.
For the parameter sets $\Omega_{ef}=\{43,71,114\}$~MHz and $T=\{1000,500,300\}$~ns, marked with stars in Fig.~2(c) of the main text, we repeat the initialization measurement 200 times to gather statistics, as depicted in Fig.~\ref{fine_fidelity}. The corresponding measured errors of $e_i$ and $e_i+e_\downarrow$ are $\{0.62\%, 0.66\%, 0.63\%\}$ and $\{1.23\%, 1.26\%, 1.18\%\}$, respectively. 

Given an initialization rate, the steady-state thermal population can be estimated using $e_{\text{steady}} = \Gamma_{\uparrow} / (\Gamma_{\uparrow} + \Gamma_\text{init})$. Here, $\Gamma_{\uparrow} = n_\text{th} /T_1$ is the excitation rate of qubit, $n_\text{th}$ is the equilibrium excited population, and $\Gamma_\text{init}$ is the initialization rate, as depicted in Fig.~\ref{simulation}(b). For $Q_A$, the thermal equilibrium population of the excited state $n_\text{th}=17.7\%$. Consequently, the theoretically predicted errors for $\Omega_{ef}= \{43,71,114\}$~MHz are $\{0.072\%, 0.042\%, 0.031\%\}$.
According to our error metrology, $e_i$ encompasses a component of the ground state's excitation error. As such, the measured initialization infidelity could serve as the upper limit for this protocol. The precision of the initialization error characterization could be potentially further refined by better readout using a Josephson parametric amplifier.

In the scenarios when the system is prepared in the second-excited state, the leakage error cannot be ignored. To quantify the leakage removal efficiency of the initialization process, we subtract the two equations in Eq.~\ref{error_func} to get 
\begin{gather}
    (\avg{\vec{r}_{g}} - \avg{\vec{r}_{e}}) / (\vec{r}_{g} - \vec{r}_{e}) = (1 - 2 e_i - e_\downarrow + P_f) \leq (1 - e_\downarrow - P_f).
\end{gather}
The inequality is equal if and only if $e_i=P_f$.
Here, we use $(\avg{\vec{r}_{g}} - \avg{\vec{r}_{e}}) / (\vec{r}_{g} - \vec{r}_{e}) + e_\downarrow$ to denote the lower bound of the leakage removal efficiency of $1-P_f$. The relevant data are plotted in the Fig.~3 of the main text.

\section{Initialization via blue sideband transition}

\begin{figure}
  \includegraphics[scale=1.0]{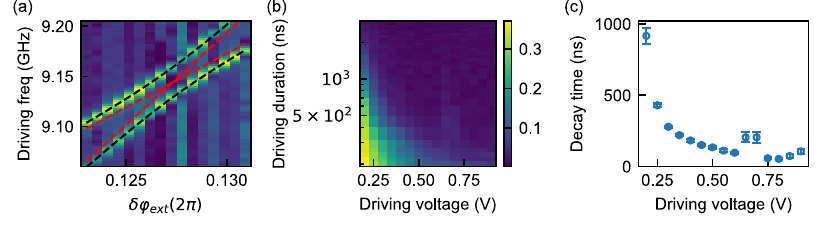}
  \caption{\label{reset_to_estate}The avoided level crossing between $|h0\rangle$ and $|e1\rangle$. The measured initialization error as a function of the driving duration and driving amplitude. (c) The extracted decay time as a function of the driving amplitude.}
\end{figure}

In addition to achieving qubit initialization via $|f0\rangle$, a comparable process can be executed by utilizing the coupling between $|h0\rangle$ and $|e1\rangle$. By substituting $|e0\rangle$, $|f0\rangle$ and $|g1\rangle$ with $|g0\rangle$, $|h0\rangle$ and $|e1\rangle$, we can facilitate a similar evolution to initialize the qubit. Employing a monochromatic drive with a frequency $\omega_r+\omega_{ge}$, a blue sideband-transition from $|g0\rangle$ to $|e1\rangle$ will occur, resetting the qubit to its first excited state $|e0\rangle$. When $|h0\rangle$ and $|e1\rangle$ are tuned into resonance, a similar level crossing can be observed in the spectral measurement, depicted in Fig.~\ref{reset_to_estate}(a). The resonance frequency and flux offset $\delta\varphi_{\text{ext}}/2\pi$ are $9.143$~GHz and $0.127$, respectively. The extracted coupling strength $g_{rh}/2\pi$ is 29.8~MHz. Compared to the initialization assisted with $|f0\rangle$, this scheme requires a higher driving frequency and a smaller flux offset away from the flux sweet spot. Same as in the previous experiment, we examine the initialization error as functions of the driving duration and amplitude at the resonance point, as shown in Fig.~\ref{reset_to_estate}(b). We can also determine the population decay time which decreases with increasing driving amplitude, as illustrated in Fig.~\ref{reset_to_estate}(c).

\bibliography{sample}